\RequirePackage{lineno}
\documentclass[prl,twocolumn,preprintnumbers,amsmath,amssymb]{revtex4}

\usepackage{graphicx}
\usepackage{lineno}
\usepackage[colorlinks,linkcolor=blue,anchorcolor=blue,citecolor=blue]{hyperref}
\begin{document}


\title{Inelastic electron tunneling in 2H-Ta$_x$Nb$_{1-x}$Se$_2$ evidenced by scanning tunneling spectroscopy}

\author{Xing-Yuan Hou$^{1,2}$,  Fan Zhang$^{2,3}$, Xin-Hai Tu$^{4,5}$  }
\author{Ya-Dong Gu$^{2,3}$, Meng-Di Zhang$^{2,3}$, Jing Gong$^{2,3}$}
\author{Yu-Bing Tu$^{1}$}
\author{Bao-Tian Wang$^{5}$}
\author{Wen-Gang Lv$^{2}$, Hong-Ming Weng$^{2,3}$, Zhi-An Ren$^{2,3}$}
\author{Gen-Fu Chen$^{2,3}$}
\author{Xiang-De Zhu$^{4}$}\email{xdzhu@hmfl.ac.cn}
\author{Ning Hao$^{4}$}\email{haon@hmfl.ac.cn }
\author{Lei Shan$^{1,2,3}$}\email{lshan@ahu.edu.cn}

\affiliation{$^1$Key Laboratory of Structure and Functional Regulation of Hybrid Materials of Ministry of Education, Institutes of Physical Science and Information Technology, Anhui University, Hefei 230601, China}
\affiliation{$^2$Beijing National Laboratory for Condensed Matter Physics, Institute of Physics, Chinese Academy of Sciences, Beijing 100190, China}
\affiliation{$^3$School of Physical Sciences, University of Chinese Academy of Sciences, Beijing 100190, China}
\affiliation{$^4$Anhui Province Key Laboratory of Condensed Matter Physics at Extreme Conditions, High Magnetic Field Laboratory, Chinese Academy of Sciences, Hefei 230031, China}
\affiliation{$^5$Institute of High Energy Physics, Chinese Academy of Sciences, Beijing 100049, China}

\date{\today}

\begin{abstract}
We report a detailed study of tunneling spectra measured on 2H-Ta$_x$Nb$_{1-x}$Se$_2$ ($x=0\sim 0.1$) single crystals using a low-temperature scanning tunneling microscope. The prominent gap-like feature unintelligible for a long time was found to be accompanied by some ``in-gap" fine structures. By investigating the second-derivative spectra and their temperature and magnetic field dependencies, we were able to prove that inelastic electron tunneling is the origin of these features and obtain the Eliashberg function of 2H-Ta$_x$Nb$_{1-x}$Se$_2$ at atomic scale, providing a potential way to study the local Eliashberg function and phonon spectra of the related transition-metal dichalcogenides.
\end{abstract}


\maketitle



As a prototype of transition-metal dichalcogenides (TMDs), 2H-NbSe$_2$ keeps attracting great interest since it provides a platform, both in bulk and in its two-dimensional form, to study the mechanism of charge-density waves (CDWs) \cite{CDWinTMD_Wilson_AiP1975, CDW_Gruner_RMP1988, NbSe2_mono_stm}, interplay between CDWs and superconductivity (SC) \cite{NbSe2_arpes_CDWboostSC_NP2007, NbSe2_CDW-SC_NC2015, NbSe2_CDW-SC_NC2018}, potential applications in electronic devices \cite{NbSe2_VdW_NP2015, NbSe2_VdW_NC2016, NbSe2_VdW_NL2017}, and so on. In recent years, there is growing consensus that it is the wave-vector-dependent strong electron-phonon coupling (EPC) that drives CDW formation \cite{NbSe2_EPC_PRL2011, NbSe2_EPC_NC2015, CDW_review}. However, unlike the typical strong EPC system of Pb \cite{Pb_Phonon1}, no characteristics of the tunneling spectra of 2H-NbSe$_2$ have been identified as its EPC function, i.e., the Eliashberg function. The dominant low-energy feature other than the superconducting gap is the kinks at approximately $\pm$35 mV, which has been observed repeatedly by scanning tunneling microscopy/spectroscopy (STM/STS) \cite{NbSe2_STM_CDWgap_Hess1991, STM_NbSe2_Hoffman_PNAS2013}. This well-known gap-like feature was initially regarded as the CDW gap \cite{NbSe2_STM_CDWgap_Hess1991}, but has not been demonstrated by subsequent measurements of angle-resolved photoelectron spectroscopy (ARPES) \cite{NbSe2_arpes_EPC_Valla_PRL2004, NbSe2_arpes_CDWgap_PRL2009}. A recent STM/STS experiment ascribed it to the combination of a 35-mV electron self-energy effect and a non-symmetric CDW gap of approximately 12 meV \cite{STM_NbSe2_Hoffman_PNAS2013}, which is supported by the later ARPES experiment \cite{NbSe2_CDW-SC_NC2015}. Nonetheless, it is confusing to note that no other phonon modes have been noticed by STS, although various modes in addition to the 35-meV one can be observed in the energy-momentum dispersions from ARPES \cite{NbSe2_ARPES_Rahn_PRB2012}.

In this work, we performed precise STM/STS measurements for 2H-Ta$_x$Nb$_{1-x}$Se$_2$ single crystals. In addition to the 35-mV gap-like feature, a series of ``in-gap" shoulders/steps were observed in all samples. By investigating the second-derivative spectra of $d^2I/dV^2$ and its temperature and magnetic field evolutions, we demonstrated that these spectroscopic features, including the long-term incomprehensible 35-mV enigma, originate from the inelastic electron tunneling process instead of a CDW gap \cite{ NbSe2_STM_CDWgap_Hess1991} or the electron self-energy effect \cite{STM_NbSe2_Hoffman_PNAS2013}. Meanwhile, a double Fermi-Dirac broadening effect was found to be a convenient way to distinguish an inelastic electron tunneling process. Then, the real-space resolved Eliashberg functions were obtained from inelastic electron tunneling spectroscopy (IETS), which might be extended to the study of other members of the TMD family.

The 2H-Ta$_x$Nb$_{1-x}$Se$_2$ single crystals studied here were grown with the iodine vapor transport method \cite{Nb1-xTaxSe2_sample1986}. The spatially resolved tunneling experiments were carried out on a home-built low-temperature scanning tunneling microscope, in which a single-crystalline sample was cold-cleaved \emph{in situ} and then inserted into the STM head immediately. Electrochemically etched tungsten tips were used for the STM/STS measurements after field-emission treatment on a piece of gold. The tunneling spectra of $dI/dV\sim V$ were obtained using a standard lock-in technique with a frequency of 373.1 Hz and a modulation amplitude of 0.1 or 0.2 mV \cite{SI}.

\begin{figure}[]
\hspace{0cm}\includegraphics[scale=0.55]{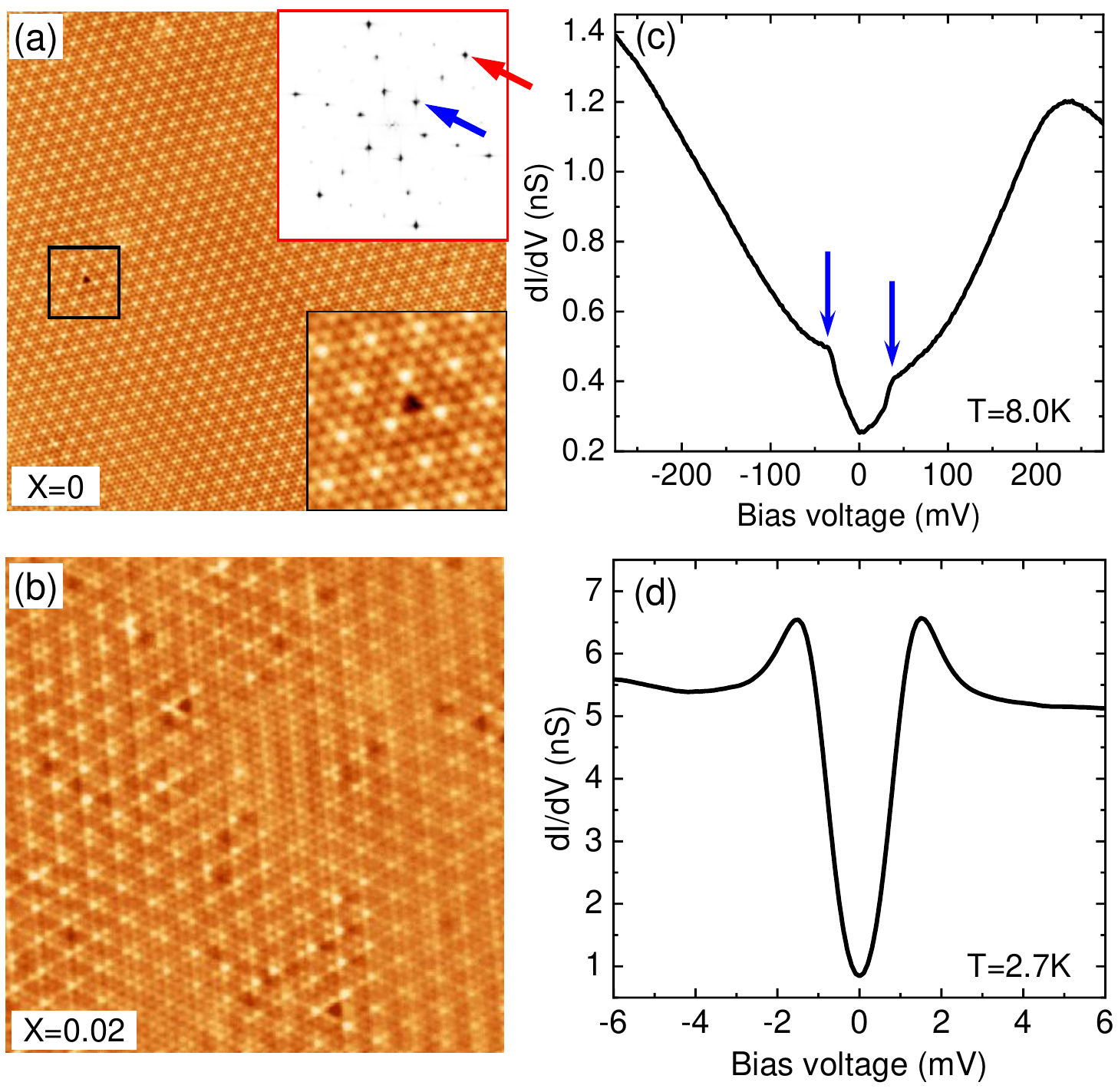}
\caption{\label{fig:fig1} (Color online) (a) 31$\times$31-nm topographic STM image of 2H-NbSe$_2$ single crystal cold-cleaved \emph{in situ}. It was taken with a sample-tip voltage of 50 mV and tunneling current of 100 pA. Inset shows the Fourier transform of the atomically resolved image; atomic peaks and CDW wave vectors are indicated by red and blue arrows, respectively. (b) 15$\times$15-nm topographic STM image of 2H-Ta$_{0.02}$Nb$_{0.98}$Se$_2$ single crystal. It was taken with a sample-tip voltage of 50 mV and tunneling current of 50 pA. (c) Large-energy-range tunneling spectrum of 2H-NbSe$_2$ measured at temperature below $T_{CDW}$. The well-known 35-mV inflexion points are marked by arrows. (d) Tunneling spectrum of 2H-NbSe$_2$ measured below $T_c$ showing a clear superconducting gap.}
\end{figure}

2H-NbSe$_2$ shows a CDW transition at $T_{CDW} \approx$ 33 K and a superconducting transition of $T_c \approx$ 7.1 K \cite{NbSe2_CDW_NS, NbSe2_CDW_NS_PRB1977}. Figure~\ref{fig:fig1}(a) shows a topographic image of 2H-NbSe$_2$, in which the atomic resolution is demonstrated by the single vacancy of an Se atom enclosed in the black box. CDW order can be clearly seen as the bright spots almost every three atoms along the sixfold- symmetric axes, and the corresponding wave vector is indicated in the inset. It is known that a moderate Ta doping into Nb sites in pristine 2H-NbSe$_2$ will impair CDW order, whereas it does not seriously change the electronic structures \cite{RennerPRL1991}. As demonstrated in Fig.~\ref{fig:fig1}(b), the long range CDW is broken into domains with nanometer scales, in some of which CDW has been strongly suppressed. Figure~\ref{fig:fig1}(c) gives a large-bias range tunneling spectrum of 2H-NbSe$_2$, where a prominent gap-like feature shows up between the inflection points at approximately $\pm$35 mV. At low temperatures below $T_c$, a superconducting gap ($\Delta\sim$1.1meV) develops as presented in Fig.~\ref{fig:fig1}(d). The 35-meV gap-like feature has long been regarded as the CDW gap \cite{NbSe2_STM_CDWgap_Hess1991}, while its equivalent in ARPES data has never been found \cite{NbSe2_arpes_EPC_Valla_PRL2004, NbSe2_arpes_CDWgap_PRL2009, NbSe2_ARPES_Rahn_PRB2012}.

\begin{figure}[]
\includegraphics[scale=0.55]{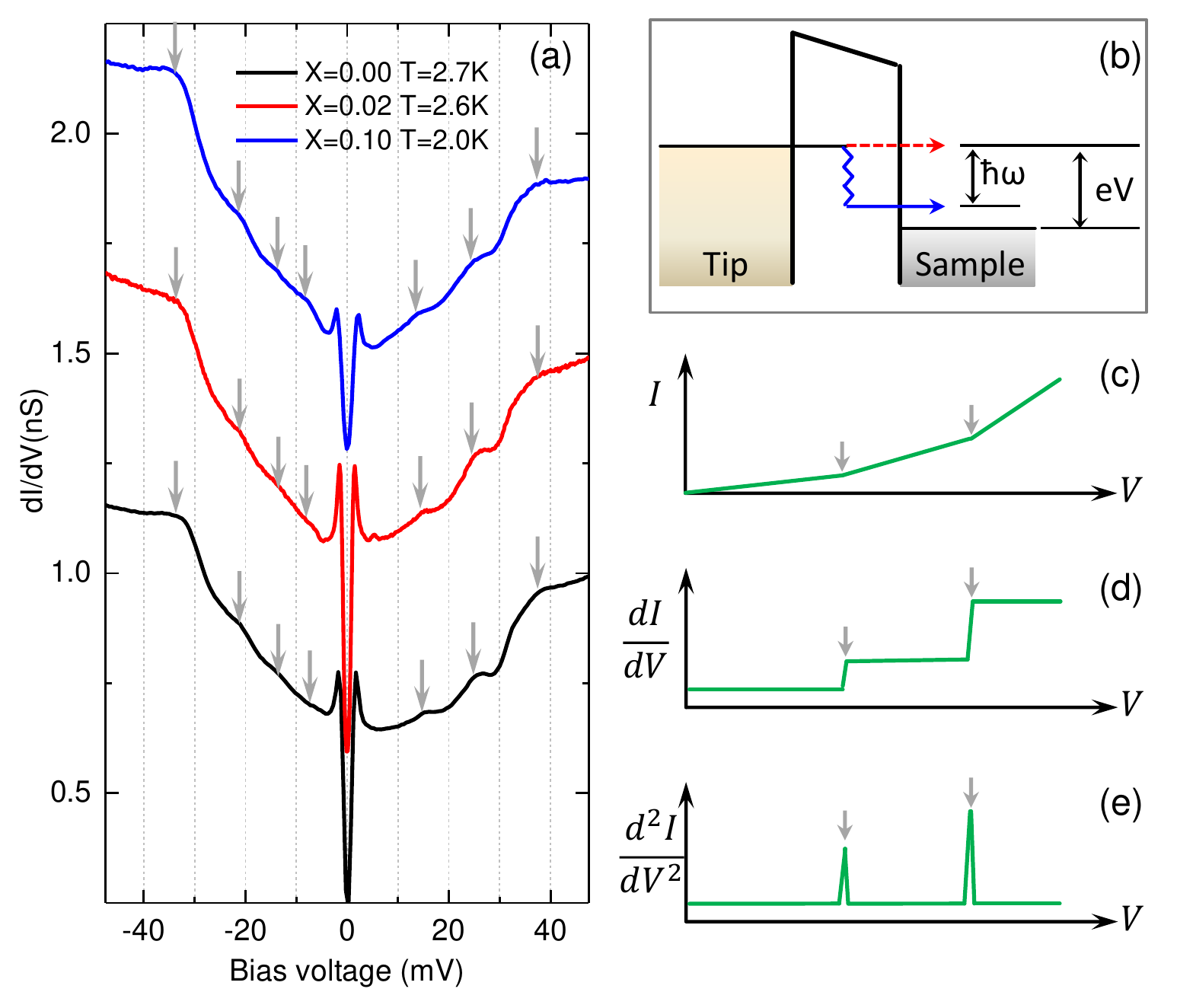}
\caption{\label{fig:fig2} (Color online) (a) Low-temperature tunneling spectra obtained on 2H-Ta$_x$Nb$_{1-x}$Se$_2$ single crystals with different doping levels ($x$). Spectra are normalized and shifted for clarity. All the features of shoulders/steps are marked by arrows. (b) Schematic of inelastic electron tunneling process taking place at a STM-based vacuum tunneling junction, in which $\hbar\omega$ is phonon energy and $eV$ the energy difference between tip and sample. (c)-(e) Illustrative inelastic electron tunneling spectra of $I-V$, $dI/dV-V$, and $d^2I/dV^2-V$, respectively, where two phonon modes are involved and indicated by arrows.}
\end{figure}

To examine the possible influence of CDWs on the tunneling spectrum, we have studied a series of 2H-Ta$_x$Nb$_{1-x}$Se$_2$ ($x = 0, 0.02, 0.1$) single crystals. As shown in Fig.~\ref{fig:fig2}(a), both the superconducting gap and the 35-mV ``gap" are very clear for all the samples. Remarkably, distinct fine structures were found inside the big ``gap" showing themselves as a series of shoulders or steps (including the 35-meV one) marked by the arrows. This is reminiscent of the inelastic electron tunneling process, the schematic of which is given in Fig.~\ref{fig:fig2}(b). In brief, when the energy of tunneling electrons reaches that of a vibrational mode in a EPC system, they could lose their energies to excite the phonon mode and open an inelastic tunneling channel \cite{review_IETS}. However, due to Fermi's golden rule, such inelastic channel is forbidden if the tunneling electron's energy is below the quantized phonon energy. In experiments, the opening of additional inelastic channels will cause slope changes of a $I-V$ curve, sharp shoulders/steps in the tunneling conductance of $dI/dV$, and corresponding peaks at each phonon energy in $d^2I/dV^2$, as depicted in Figs.~\ref{fig:fig2}(c)-(e). Actually, $d^2I/dV^2$ versus $V$ directly reflects the effective EPC spectrum $\alpha^2F(\omega)$, i.e., the Eliashberg function \cite{taylor1992inelastic,Pb_IETS_PRL2015}. Here, $F(\omega)$ is the phonon density of states (DOS) and $\alpha$ the energy-dependent EPC strength. This method, known as IETS, could be utilized to reveal the vibrational spectra of molecules at the interface of a tunneling junction \cite{IETS_Ho_Science1998, IETS_review_Ho, IETS_review_JOP2007, review_IETS} or, in a few cases, to obtain the collective vibrations or Eliashberg functions in solids \cite{graphite_IETS_PRB2004, Au_IETS_2008, Pb_IETS_PRL2012, Pb_IETS_PRL2015, Pb_IETS_PRB2016}.

\begin{figure}[]
\includegraphics[scale=0.55]{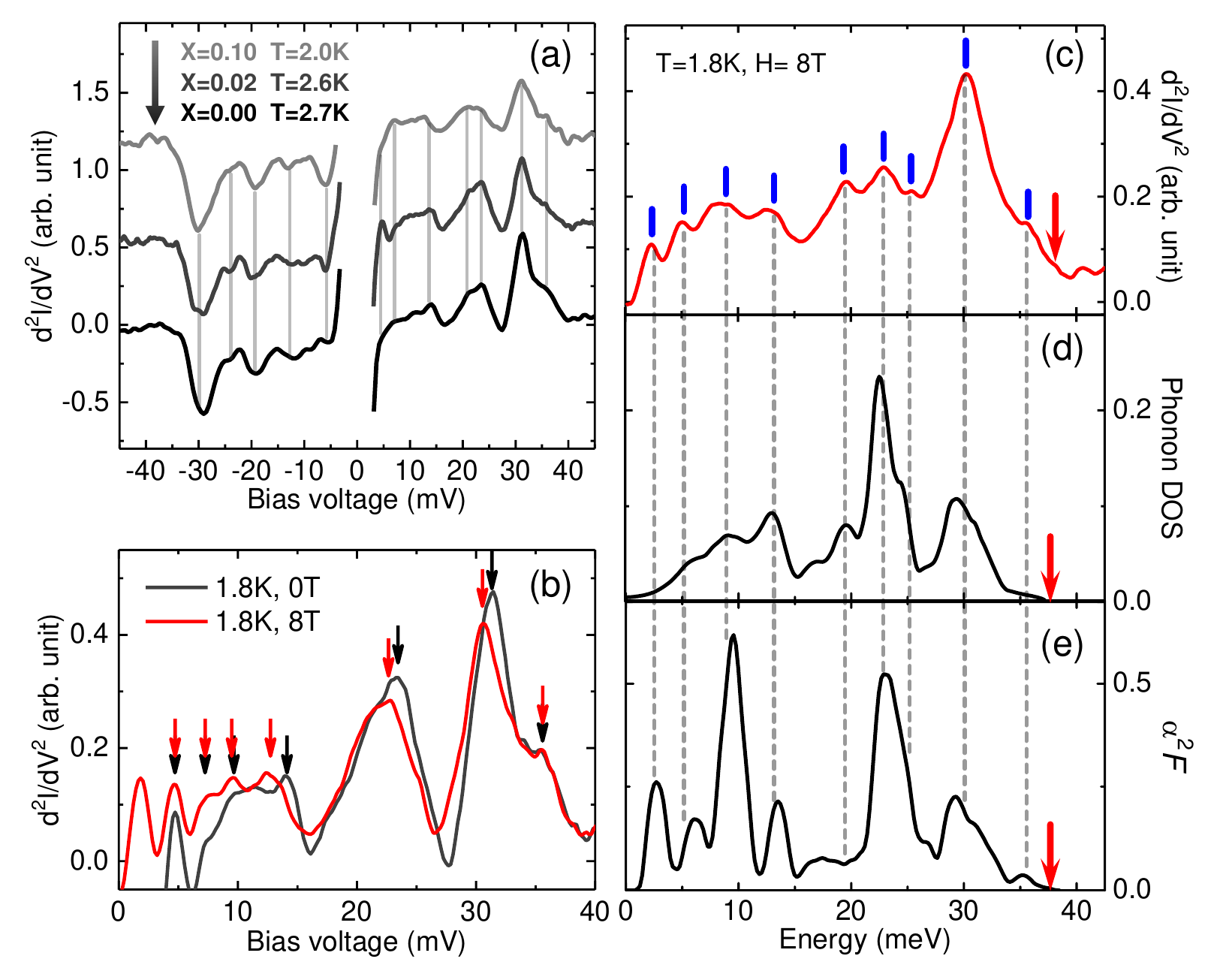}
\caption{\label{fig:fig3} (Color online) (a) Spectra of $d^2I/dV^2$ derived from data in Fig.~\ref{fig:fig2}(a). The most prominent phonon modes are marked by vertical lines, almost irrelevant regarding the doping levels. Differences between the features of the positive- and negative-bias parts are most likely due to particle-hole asymmetry in the electron DOS. (b) Comparison of low-temperature spectra measured in the superconducting state and normal state with $H$=0T and 8T, respectively. Data were taken on sample with $x=0.02$. (c)-(e) Comparison between experimental data and calculated phonon DOS and Eliashberg function $\alpha^2F(\omega)$. Positive- and negative-bias parts of measured spectra have been averaged and calculated curves have been smoothed to match energetic resolution of measurements.   }
\end{figure}

Except for the non-symmetric background of the electron DOS and superconducting gap, all the spectral features shown in Fig.~\ref{fig:fig2}(a) look qualitatively similar to that of Fig.~\ref{fig:fig2}(d). Figure~\ref{fig:fig3}(a) displays the $d^2I/dV^2$ spectra derived from the data in Fig.~\ref{fig:fig2}(a), showing similar shapes for all the doping levels and excluding the CDW order as the major origin of the observed fine spectral structures and the enigmatic 35-mV ``gap". In contrast, the Eliashberg function was expected to be not obviously dependent on the Ta doping up to a content of 10\% according to our first-principles calculations \cite{SI}. Furthermore, as shown in Fig.~\ref{fig:fig3}(b), when superconductivity is suppressed by a magnetic field and hence the superconducting gap is closed, some peaks shift slightly towards lower energies while some others do not. This is in good agreement with the momentum-dependent EPC and superconducting gap $\Delta(k)$ in 2H-NbSe$_2$ \cite{NbSe2_arpes_EPC_Valla_PRL2004, NbSe2_ARPES_Rahn_PRB2012}. In the framework of IETS, a peak related to a specific phonon $\hbar\omega$ in the normal state will shift to $\hbar\omega+\Delta(k)$ in the superconducting state. Thus, the IETS peaks coupled to different parts of the Fermi surfaces will shift to different energy scales determined by the momentum dependence of $\Delta(k)$. As shown in Figs.~\ref{fig:fig3}(c)-(e), both the phonon DOS and Eliashberg function of 2H-NbSe$_2$ were calculated \cite{SI} in comparison with the experimental data. Almost all the characteristic peaks marked in Fig.~\ref{fig:fig3}(c) have their counterparts in the calculated phonon DOS or $\alpha^2F(\omega)$ as given in Figs.~\ref{fig:fig3}(d) and (e). Furthermore, the upper boundary of the phonon DOS or $\alpha^2F(\omega)$ at approximately 37 meV is very close to the value of 38 meV in $d^2I/dV^2$, as denoted by the arrows.

\begin{figure}[]
\includegraphics[scale=0.55]{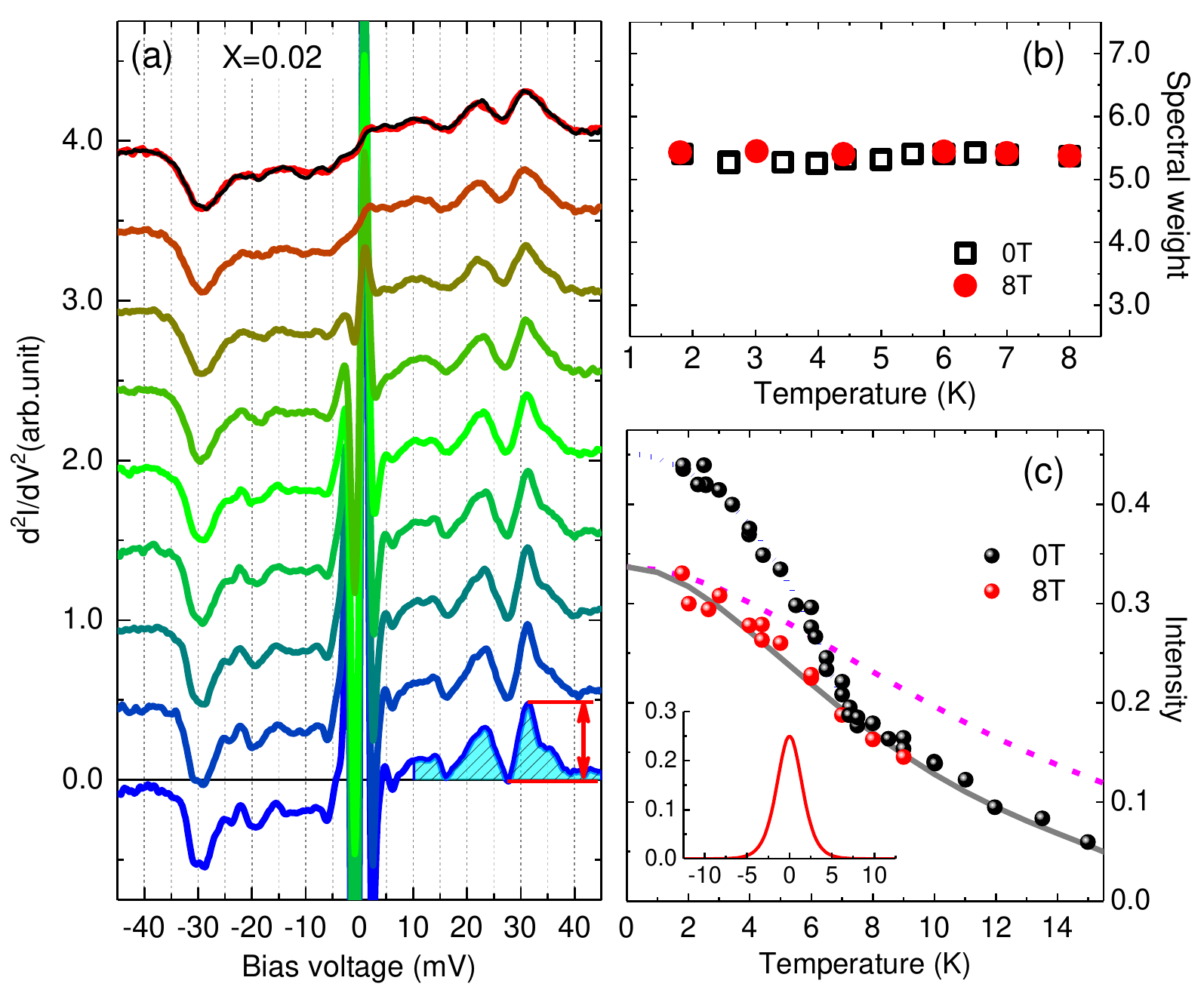}
\caption{\label{fig:fig4} (Color online)  (a) Temperature evolution of $d^2I/dV^2$ spectra measured at 1.8 $\sim$ 8 K (from bottom up). Top curve of 8 K (red) is overlapped with that taken at same temperature while under 8 T (black). (b) Temperature dependence of spectral weight integrated from 10 to 45 mV, as illustrated by the shadow in (a). (c) Temperature dependence of a peak intensity indicated by two-way arrow in (a). The two sets of data obtained at 0 and 8 T begin to diverge at $T_c$ with decreasing temperature. Dashed and solid lines are simulations considering a single Fermi-Dirac broadening and a double one, respectively. Inset illustrates the Fermi-Dirac broadening function. }
\end{figure}

The identification of the Eliashberg function here elucidates the mechanism of the 35-mV enigma as an inelastic electron tunneling process. This provides a good opportunity to acquire knowledge of the IETS in a superconducting system. To this end, we have measured the temperature and magnetic field dependencies of IETS. As exemplified by Fig.~\ref{fig:fig4}(a), the peaks in $d^2I/dV^2$ decay gradually with increasing temperature, but survive well above $T_c$. In the normal state at $T$ = 8 K, the data of zero field look the same as that of a 8-T field, ruling out any form of magnetic effect on EPC. In Fig.~\ref{fig:fig4}(b), the spectral weights integrated between 10 meV (selected to exclude the influence of superconducting gap at lower bias) and 45 meV are plotted against temperature for both zero field and 8 T. No obvious temperature dependence can be seen, indicating a nearly constant Eliashberg function in the studied temperature range. Thus, the observed peak decaying with temperature should originate from the thermal broadening effect. To confirm this judgement, we present in Fig.~\ref{fig:fig4}(c) the temperature dependence of the peak intensity defined as the drop between the highest peak and the dip beside it as shown in Fig.~\ref{fig:fig4}(a). It was found that the peak intensity enhances continuously with decreasing temperature and an extra enhancement occurs in the superconducting state compared with the normal state.

We have tried to build a simple model to quantitatively explain the normal-state data, in which inelastic tunneling is considered as a two-step process and the attenuation of peak intensity is due to Fermi-Dirac (thermal) broadening. For an ordinary tunneling process at a finite temperature, the thermal broadening of the tunneling spectrum is rooted in the Fermi-Dirac distributions of the occupied electron DOS at both banks of the tunneling junction. If the electron DOS stays the same in the studied temperature range, the broadening effect is equivalent to a convolution of the zero-temperature spectrum with a temperature-dependent Fermi-Dirac broadening function as illustrated in the inset of Fig.~\ref{fig:fig4}(c). In contrast, inelastic tunneling includes two steps, i.e., the ordinary tunneling and a subsequent transition to a lower energy by exciting a phonon. Since the Fermi-Dirac broadening effect takes effect in the second process as well, a double Fermi-Dirac broadening should be considered in our calculations \cite{SI}. The good agreement between this simple model [see the solid line in Fig.~\ref{fig:fig4}(c)] and experimental data further support that the fine spectral structures of 2H-Ta$_x$Nb$_{1-x}$Se$_2$ are indeed related to the inelastic electron tunneling process. The line calculated with single Fermi-Dirac broadening is also presented, significantly deviating from the experimental data. Therefore, the constructed model provides a convenient way to distinguish IETS from other mechanisms, such as the self-energy effect in which single Fermi-Dirac broadening is involved.

According to the IETS mechanism, it is easy to understand the above-mentioned additional enhancement of peak intensity in the superconducting state. The peak intensity at a particular phonon energy is determined by the phonon DOS at this energy and the electron DOS first involved in the inelastic channel. Since quasiparticle DOS peaks (coherence peaks) develop at the gap edges below $T_c$, the IETS peak will become stronger accordingly, in comparison with the normal state. With decreasing temperature, the coherence peaks are intensified continuously, resulting in an increasingly more prominent IETS peak. Therefore, our experiments indicate that such electron DOS redistribution can be used to enhance spectral intensity and discern the Eliashberg function more easily.

\begin{figure}[]
\includegraphics[scale=0.52]{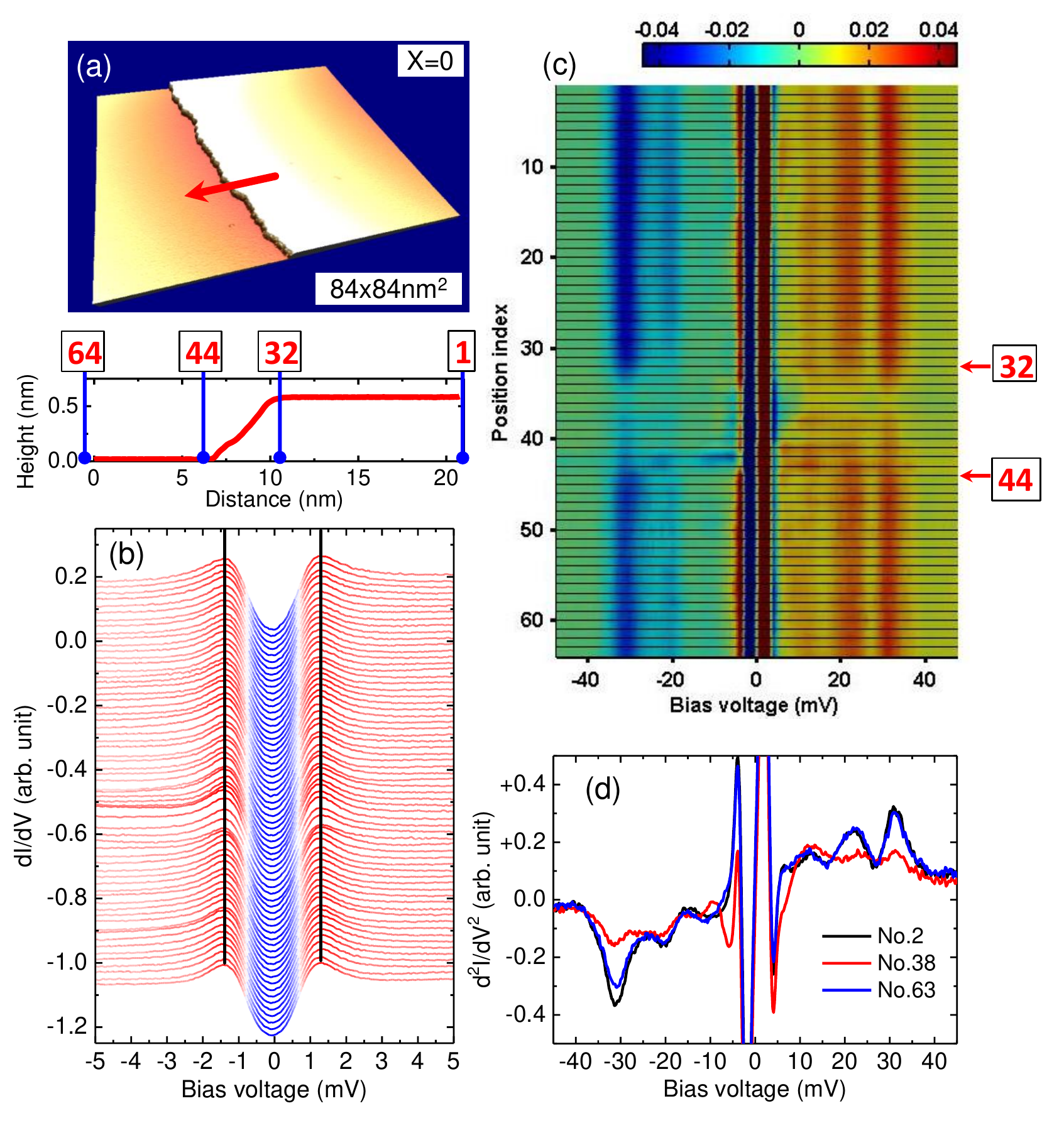}
\caption{\label{fig:fig5} (Color online)  (a) Large-scale topographic image containing a step across the entire region. A trajectory of 21 nm is indicated by red arrow with its section profile plotted under the image, along which both superconducting spectra and $d^2I/dV^2$ spectra were taken sequentially at 64 evenly spaced points as given in (b) and (c), respectively. All data were obtained at 2.4 K. (d) Comparison of data taken at the step and far from it. }
\end{figure}

Using STM/STS offers the possibility to measure IETS at the atomic scale. However, for the moment, this methodology has been limited to a few solid materials, such as graphite/graphene \cite{graphite_IETS_PRB2004, graphene_NP2008_ZhangYuanbo, graphene_2010, graphene_IETS_PRL2015}, copper \cite{Au_IETS_2008, Cu_IETS_PRB2016}, gold \cite{Au_IETS_2008}, and the only superconducting material of lead \cite{Pb_IETS_PRL2012, Pb_IETS_PRL2015, Pb_IETS_PRB2016}. The observation in this work extends the field of research to the widely concerned TMDs, an easily tailored system suitable to further study or manipulate EPC. For example, we could examine the influence of a structural defect on the Eliashberg function and superconductivity of 2H-NbSe$_2$ simultaneously. Figure~\ref{fig:fig5}(b) shows a series of $dI/dV$ spectra taken sequentially at 64 points along a line across a half-unit-cell step displayed in Fig.~\ref{fig:fig5}(a). The superconducting gap remains almost constant along the trajectory, whereas the $d^2I/dV^2$ spectrum of Fig.~\ref{fig:fig5}(c) shows a remarkable change in the step region of points 32-44. As presented in Fig.~\ref{fig:fig5}(d), the spectral weight at the step is seriously suppressed above 15 mV compared with those two spectra taken near the opposite ends of the trace, which should be related to the abrupt changes of either electron/phonon DOS or EPC. Therefore, the spatially resolved spectra given in Fig.~\ref{fig:fig5} clearly demonstrate the immunity of the bulk Bardeen-Cooper-Schrieffer superconductivity to a local variation of the Eliashberg function. The situation is expected to be different in a size-limited system (say, a few-layered one) and thus deserves more attention in the future.\nocite{giannozzi2009quantum, hohenberg1964inhomogeneous, kohn1965self, baroni2001phonons, perdew1996generalized, blochl1994projector, kresse1999ultrasoft, PhysRevLett.115.136402, PhysRevB.76.125112}.

In summary, tunneling spectra of 2H-Ta$_x$Nb$_{1-x}$Se$_2$ single crystals have been measured at low temperatures and high magnetic fields. Using first-principles calculations and a simple two-step tunneling model, we were able to relate the derived spectra of $d^2I/dV^2$ to the inelastic electron tunneling process and hence the Eliashberg functions. These results not only resolve the long-standing enigma of the 35-mV ``gap"-like feature observed in 2H-NbSe$_2$, but also pave a way to explore the spatially dependent electron-phonon coupling in the widely concerned transition-metal dichalcogenides.


\begin{acknowledgments}

We thank T. Xiang, C. Ren, and K. Jin for fruitful discussions. This work was financially supported by the National Key R\&D Program of China (Grant No. 2017YFA0403502, 2017YFA0302904, 2018YFA0305602, 2017YFA0303201), National Natural Science Foundation of China (Grant No. 11574372, 11322432, 11804379, 11674331), ``100 Talents Project" of the Chinese Academy of Sciences,  CASHIPS Director's Fund (Grant No. BJPY2019B03), Youth Innovation Promotion Association of CAS (Grant No. 2017483), and The Recruitment Program for Leading Talent Team of Anhui Province  (2019-16). A portion of this work was supported by the High Magnetic Field Laboratory of Anhui Province.

\end{acknowledgments}

\bibliography{IETS}

\end{document}